\documentclass[pra,showpacs,showkeys]{revtex4}
\usepackage{graphicx,amsmath,amsfonts,amssymb}
\usepackage{times}
\begin{document}
\title{\Large Remote information concentration and multipartite entanglement in multilevel systems}
\author{Xin-Wen Wang$^{1,}$\footnote{xwwang@mail.bnu.edu.cn}, Deng-Yu
Zhang$^{1,}$\footnote{dyzhang672@163.com}, Guo-Jian
Yang$^{2,}$\footnote{yanggj@bnu.edu.cn}, Shi-Qing Tang$^1$, and
Li-Jun Xie$^1$}
 \affiliation{$^1$Department of Physics and Electronic Information,
  Hengyang Normal University, Hengyang 421008, People's Republic of China\\
  $^2$Department of Physics, Beijing Normal University, Beijing 100875, People's Republic of China}

\begin{abstract}
Remote information concentration (RIC) in $d$-level systems (qudits)
is studied. It is shown that the quantum information initially
distributed in three spatially separated qudits can be remotely and
deterministically concentrated to a single qudit via an entangled
channel without performing any global operations. The entangled
channel can be different types of genuine multipartite pure
entangled states which are inequivalent under local operations and
classical communication. The entangled channel can also be a mixed
entangled state, even a bound entangled state which has a similar
form to the Smolin state, but has different features from the Smolin
state. A common feature of all these pure and mixed entangled states
is found, i.e., they have $d^2$ common commuting stabilizers. The
differences of qudit-RIC and qubit-RIC ($d=2$) are also analyzed.

\end{abstract}



 \maketitle

 \section{introduction}
Although an unknown quantum state cannot be perfectly copied
\cite{299N802,92PLA271}, quantum cloning, functioning as copying
approximately quantum states as well as possible, has attracted
considerable attention \cite{77RMP125} since Bu\v{z}ek and Hillery
\cite{54PRA1844} first introduced such a concept, due to its
potential applications in quantum information science (see,
e.g.,\cite{62PRA022301,73PRA032304,95PRL090504,63PRA042308}).
Although the fidelities of clones relative to the original state are
less than one, the quantum information of the input system is not
degraded but only distributed into a larger quantum system. That is,
the quantum cloning process can be regarded as the distribution of
quantum information from an initial system to final ones. Thus,
quantum cloning combined with other quantum information processing
(QIP) tasks may have potential applications in multiparty quantum
communication and distributed quantum computation. This leads to the
advent of the concept of quantum telecloning
\cite{59PRA156,61PRA032311,67PRA012323}, which is the combination of
quantum cloning and quantum teleportation \cite{70PRL1895}.
Telecloning functions as transmitting many copies of an unknown
quantum state of the input system to many distant quantum systems,
i.e., realizing one-to-many remote cloning, via previously shared
multipartite entangled states. As the reverse process of
telecloning, remote information concentration (RIC) was also
presented by Murao and Vedral \cite{86PRL352}. They demonstrated
that the quantum information originally distributed into three
spatially separated qubits from a single qubit can be remotely
concentrated back to a single qubit via a four-qubit unlockable
bound entangled state \cite{63PRA032306,80PRL5239} without
performing any global operations. Telecloning and concentrating
processes could be regarded as, respectively, remote information
depositing and withdrawing, or remote information encoding and
decoding, which is expected to find useful applications in
network-based QIP \cite{86PRL352}. Yu \emph{et al.}
\cite{68PRA024303} showed that a four-qubit GHZ state can also be
used to implement three-to-one RIC. Not long before, RIC was
generalized to the $N\rightarrow 1$ case in two-level systems
\cite{73PRA012318,76PRA032311}.

In recent years, encoding and manipulating quantum information with
high-dimensional systems, or qudits, instead of two-state systems,
or qubits, has attracted considerable attention. This is due to the
fact that significant fundamental and practical advantages can be
gained by employing high-dimensional quantum states. For instance,
higher-dimensional entangled states exhibit stronger violation of
local realism \cite{85PRL4418} and can lower the detection
efficiencies required for closing the detection loophole in Bell
tests \cite{104PRL060401}, higher-dimensional states are more robust
against isotropic noise \cite{90PRL167906}, qudit-based quantum
cryptographic protocols may enhance the security against
eavesdropping attacks \cite{88PRL127902}, qudits can simplify
quantum logic \cite{5NP134} and have higher capacity to carry
information, and so on.

In this paper, we investigate RIC for $d$-level ($d\geqslant 2$)
quantum systems, called qudits for short (when $d=2$, they reduce to
qubits). It will be shown that the quantum information originally
distributed into three spatially separated qudits from a single
qudit by the telecloning procedure can be remotely concentrated back
to a single qudit via a previously shared entangled channel assisted
by local operations and classical communication (LOCC). The
entangled channel can be mixed entangled states as well as pure
ones. All these entangled states have $d^2$ common commuting
stabilizers. We also show that there are minor constraints on the
distribution of the general entangled channel, in contrast to
qubit-RIC which has no constraint on the distribution of the
entangled channel.

It can be seen that entanglement, a very important physical resource
for QIP, plays an essential role in quantum cloning, telecloning,
and RIC. Quantum cloning is in fact creating entanglement among the
involved quantum systems, and the fidelities of clones are
inherently linked with the entanglement among them. Both telecloning
and RIC protocols need special structure of entangled states acting
as the quantum channel. In a word, all the aforementioned tasks
cannot be achieved without entanglement.

On the other hand, the quantum tasks mentioned above can reveal some
peculiar entanglement characteristics
\cite{86PRL352,68PRA024303,73PRA012318,76PRA032311,81PRA032323,79PRA064306,79PRA062315},
in addition to their practical applications. In this paper, we
reveal other interesting phenomena that appear in the RIC. A
counterintuitive phenomenon is that inequivalent \emph{genuine}
four-partite pure entangled states, i.e., they cannot be transformed
into each other by LOCC, can implement deterministically a same
\emph{multiparty} QIP task, three-to-one RIC. Another phenomenon is
that a single asymmetric unlockable bound entangled state can be
competent for implementing RIC in multilevel systems. Such a
multilevel bound entangled state has a similar form to the Smolin
bound entangled state \cite{63PRA032306} (a four-qubit unlockable
bound entangled state), but has some different features from the
Smolin state.

\section{protocols for remote information concentration via different types of entanglement}
Before describing our RIC protocols, we briefly summarize the
forward process, telecloning. We focus on the $1\rightarrow 2$
universal telecloning in $d$-level systems and its reverse in this
paper. Such a telecloning scheme \cite{67PRA012323} allows direct
distribution of optimal clones from a single original qudit state
\begin{equation}
\label{single}
|\varphi\rangle_t=\sum_{j=0}^{d-1}x_j|j\rangle_t
\end{equation}
($\sum_{j=0}^{d-1}|x_j|^2=1$) to two spatially separated parties
(Bob and Charlie) with only LOCC. The quantum channel is a
four-qudit entangled state
\begin{equation}
\label{telecloning}
   |\Phi\rangle_{t'12a}=\frac{1}{\sqrt{d}}\sum\limits_{j=0}^{d-1}|j\rangle_{t'}|\phi_j\rangle_{12a},
\end{equation}
where
\begin{equation}
  |\phi_j\rangle_{12a}=Q\left[|j\rangle_1|j\rangle_2|j\rangle_a
  +\sum\limits_{r=1}^{d-1}(p|j\rangle_1|\overline{j+r}\rangle_2+q|\overline{j+r}\rangle_1|j\rangle_2)|\overline{j+r}\rangle_a\right]
\end{equation}
with $Q=1/\sqrt{1+(d-1)(p^2+q^2)}$, $p+q=1$, and
$\overline{j+r}=j+r$ modulo $d$. Here qudit $t'$ is an input port of
the distributor, qudit $a$ is an output port for the ancilla held by
Alice, and qudits 1 and 2 are output ports for the clones held,
respectively, by Bob and Charlie (throughout the paper, if
necessary, the subscripts outside the kets or of the operators
denote the qudit index). The distributor performs a generalized (or
qudit) Bell-basis [see Eq.~(\ref{Bell})] measurement (GBM) on qudits
$t$ and $t'$. Depending on the distributor's measurement outcome,
Alice, Bob, and Charlie perform local operations on the qudits they
hold, and obtain the cloning state of $|\varphi\rangle$ represented
by the three-qudit state
\begin{equation}
\label{clone}
  |\psi\rangle_{12a}=\sum\limits_{j=0}^{d-1}x_j|\phi_j\rangle_{12a}.
\end{equation}
The aforementioned generalized ($d$-level) Bell-basis is given by
\begin{eqnarray}
\label{Bell}
  &&|B^{0,0}\rangle=\frac{1}{\sqrt{d}}\sum\limits_{j=0}^{d-1}|j\rangle|j\rangle,\nonumber\\
  &&|B^{m,n}\rangle=I\otimes U^{m,n}|B^{0,0}\rangle,\nonumber\\
  && U^{m,n}=\sum\limits_{k=0}^{d-1}\omega^{km}|\overline{k+n}\rangle\langle k|
\end{eqnarray}
for $0\leqslant m,n\leqslant d-1$, where $\omega=e^{2\pi i/d}$. In
the telecloning scheme above, when $p=q=1/2$, the cloning is
symmetric (two clones have the same fidelity) \cite{81PRL5003}, and
otherwise, it is asymmetric (two clones have different fidelities)
\cite{47JMO187}.

Using the equality
\begin{equation}
\label{equality}
  |j\rangle|k\rangle=\frac{1}{\sqrt{d}}\sum\limits_{r=0}^{d-1}\omega^{-jr}|B^{r,\overline{k-j}}\rangle~~~~(0\leqslant j,k\leqslant d-1)
\end{equation}
with $\overline{k-j}=k-j+d$ modulo $d$, we can rewrite the cloning
state of Eq.~(\ref{clone}) as
\begin{eqnarray}
\label{clone1}
 |\psi\rangle_{12a}&=&\alpha|B^{0,0}\rangle_{1a}|\varphi\rangle_2
 +\beta\sum\limits_{m=1}^{d-1}|B^{m,0}\rangle_{1a} U^{\overline{-m},0}_2|\varphi\rangle_2\nonumber\\
 &&+\gamma\sum\limits_{m=0,n=1}^{d-1}|B^{m,n}\rangle_{1a} U^{\overline{-m},n}_2|\varphi\rangle_2,
\end{eqnarray}
where
\begin{eqnarray}
 && \alpha=\frac{Q[1+(d-1)p]}{\sqrt{d}},\nonumber\\
 && \beta=\frac{Q(1-p)}{\sqrt{d}},~~~~\gamma=\frac{Qq}{\sqrt{d}}.
\end{eqnarray}
 Because of the permutability of qudits 1 and 2, the cloning
state can also be expressed as
\begin{eqnarray}
\label{clone2}
 |\psi\rangle_{12a}&=&\alpha|B^{0,0}\rangle_{2a}|\varphi\rangle_1
 +\beta\sum\limits_{m=1}^{d-1}|B^{m,0}\rangle_{2a} U^{\overline{-m},0}_1|\varphi\rangle_1\nonumber\\
 &&+\gamma\sum\limits_{m=0,n=1}^{d-1}|B^{m,n}\rangle_{2a} U^{\overline{-m},n}_1|\varphi\rangle_1.
\end{eqnarray}
When $d=2$, the results reduce to that for qubits. In other words,
the formulas of Eqs.~(\ref{clone1}) and (\ref{clone2}) can be
directly generalized from qubits to qudits. However,
\begin{equation}
\label{clone3}
  |\psi\rangle_{12a}\neq \sum\limits_{m,n=0}^{d-1}C_{mn}|B^{m,n}\rangle_{12} U^{\overline{-m},n}_a|\varphi\rangle_a
\end{equation}
for $d>2$, which can also be verified by the equality of
Eq.~(\ref{equality}). That is, the formulation of Eq.~(\ref{clone3})
cannot be generalized from qubits to qudits. Such a minor difference
will lead to the results of RIC for qudits and qubits also having
differences. Particularly, there are minor constraints on the
distribution of the general entangled channel for qudit-RIC, but
none for qubit-RIC.

Now we present our RIC schemes, the reverse of the aforementioned
universal $1\rightarrow 2$ telecloning in $d$-level systems, that
is, concentrating the information initially distributed in three
spatially separated qudits $a$, 1, and 2 (held by Alice, Bob, and
Charlie, respectively) to a single remote qudit $6$ (held by Diana)
with only LOCC: $|\psi\rangle_{12a}\rightarrow|\varphi\rangle_6$. We
first consider employing the following four-qudit pure entangled
state as the quantum channel:
\begin{equation}
\label{channelg}
  |\Psi^g\rangle_{3456}=\sum\limits_{m',n'=0}^{d-1}C_{m'n'}|B^{m',n'}\rangle_{34}|B^{\overline{u-m'},\overline{v-n'}}\rangle_{56},
\end{equation}
where $u$ and $v$ are two arbitrarily given nonnegative integers
that are less than $d$, and $C_{m'n'}$ are normalization
coefficients satisfying $\sum_{m',n'=0}^{d-1}|C_{m'n'}|^2=1$. We
first assume that qudits 3, 4, and 5 belong to Alice, Bob, and
Charlie, respectively. According to Eqs.~(\ref{clone1}) and
(\ref{channelg}), the state of the whole system
$|\Omega\rangle_{12a3456}=|\psi\rangle_{12a}|\Psi^g\rangle_{3456}$
is given by
\begin{eqnarray}
\label{whole}
 |\Omega\rangle_{12a3456}&=&\alpha\sum\limits_{m',n'=0}^{d-1}C_{m'n'}|B^{0,0}\rangle_{1a}|B^{m',n'}\rangle_{34}|\varphi\rangle_2|B^{\overline{u-m'},\overline{v-n'}}\rangle_{56}\nonumber\\
 &&+\beta\sum\limits_{\stackrel{m',n'=0}{m=1}}^{d-1}C_{m'n'}|B^{m,0}\rangle_{1a}|B^{m',n'}\rangle_{34} U^{\overline{-m},0}_2|\varphi\rangle_2|B^{\overline{u-m'},\overline{v-n'}}\rangle_{56}\nonumber\\
 &&+\gamma\sum\limits_{\stackrel{m',n'=0}{m=0,n=1}}^{d-1}C_{m'n'}|B^{m,n}\rangle_{1a}|B^{m',n'}\rangle_{34} U^{\overline{-m},n}_2|\varphi\rangle_2|B^{\overline{u-m'},\overline{v-n'}}\rangle_{56}.
\end{eqnarray}
With the equality of Eq.~(\ref{equality}), we can obtain an equality
on entanglement swapping
\begin{equation}
\label{swapping}
|B^{m,n}\rangle_{XY}|B^{m',n'}\rangle_{X'Y'}=\frac{1}{d}\sum\limits_{m'',n''=0}^{d-1}
 \omega^{m''n''}|B^{\overline{m+m''},\overline{n'+n''}}\rangle_{XY'}|B^{\overline{m'-m''},\overline{n-n''}}\rangle_{X'Y}.
\end{equation}
Using Eq.~(\ref{swapping}), the global state
$|\Omega\rangle_{12a3456}$ can be rewritten as
\begin{eqnarray}
\label{whole1}
 |\Omega\rangle_{12a3456}&=&\frac{\alpha}{d}\sum\limits_{\stackrel{m',n'=0}{m'',n''=0}}^{d-1}\omega^{m''n''}C_{m'n'}|B^{m'',\overline{n'+n''}}\rangle_{14}
  |B^{\overline{m'-m''},\overline{-n''}}\rangle_{3a}|\varphi\rangle_2|B^{\overline{u-m'},\overline{v-n'}}\rangle_{56}\nonumber\\
 &&+\frac{\beta}{d}\sum\limits_{\stackrel{m',n'=0}{\stackrel{m'',n''=0}{m=1}}}^{d-1}\omega^{m''n''}C_{m'n'}|B^{\overline{m+m''},\overline{n'+n''}}\rangle_{14}
  |B^{\overline{m'-m''},\overline{-n''}}\rangle_{3a} U^{\overline{-m},0}_2|\varphi\rangle_2|B^{\overline{u-m'},\overline{v-n'}}\rangle_{56}\nonumber\\
 &&+\frac{\gamma}{d}\sum\limits_{\stackrel{m',n'=0}{\stackrel{m'',n''=0}{m=0,n=1}}}^{d-1}\omega^{m''n''}C_{m'n'}|B^{\overline{m+m''},\overline{n'+n''}}\rangle_{14}
  |B^{\overline{m'-m''},\overline{n-n''}}\rangle_{3a} U^{\overline{-m},n}_2|\varphi\rangle_2|B^{\overline{u-m'},\overline{v-n'}}\rangle_{56}.\nonumber\\
\end{eqnarray}
The procedure of the RIC is as follows. (S1) Alice, Bob, and Charlie
perform GBMs on the qudit-pairs ($3,a$), ($1,4$), and ($2,5$),
respectively. (S2) Each party tells Diana the measurement outcome by
sending $2\log d$ bits of classical information. (S3) Diana performs
the conditional local operation on qudit 6. A schematic picture of
this protocol is shown in Fig.~1.

In (S1), the GBMs of Alice, Bob, and Charlie are independent, and
thus the sequence can be arbitrary. For clarity, we here assume that
Alice and Bob perform the GBMs before Charlie. For the outcomes
$(\overline{m'-m''},\overline{n-n''})$ and
$(\overline{m+m''},\overline{n'+n''})$, we obtain the digital values
$u'=\overline{m+m'}$ and $v'=\overline{n+n'}$. Then qudits 2, 5, and
6 are projected in the state
$U^{\overline{-m},n}_2|\varphi\rangle_2|B^{\overline{u-m'},\overline{v-n'}}\rangle_{56}$,
which can be rewritten as
\begin{eqnarray}
\label{teleportation}
 U^{\overline{-m},n}_2|\varphi\rangle_2|B^{\overline{u-m'},\overline{v-n'}}\rangle_{56}
   &=& \frac{1}{d}U^{\overline{-m},n}_2\sum\limits_{m''',n'''=0}^{d-1}U^{m''',n'''}_5
       |B^{\overline{u-m'},\overline{v-n'}}\rangle_{25}U^{\overline{-m'''},n'''}_6 |\varphi\rangle_6\nonumber\\
   &=& \frac{1}{d}\sum\limits_{m''',n'''=0}^{d-1}\omega^{n(u'-u)+(v-v')m'''}
       |B^{\overline{m'''+u-u'},\overline{n'''+v-v'}}\rangle_{25}U^{\overline{-m'''},n'''}_6 |\varphi\rangle_6.\nonumber\\
\end{eqnarray}
 Next Charlie performs a GBM
on qudits 2 and 5, which can be regarded as being equivalent to
Charlie and Diana together performing the teleportation protocol
with a local error-correction operation on qudit 6. Assume that the
measurement outcome is
$(u''=\overline{m'''+u-u'},v''=\overline{n'''+v-v'})$, and qudit 6
is projected in the state $U^{\overline{-m'''},n'''}_6
|\varphi\rangle_6$. After receiving all the measurement outcomes
sending from the other three parties, Diana can deduce the digital
values $m'''=\overline{u''+u'-u}$ and $n'''=\overline{v''+v'-v}$.
Then, Diana performs the local operation
$(U_6^{\overline{-m'''},n'''})^+=\omega^{-m'''n'''}U_6^{m''',\overline{-n'''}}$
and obtains the state $|\varphi\rangle_6$. As a consequence, the
information initially distributed in three spatially separated
qudits is now remotely concentrated in a single qudit.

If qudit 4 is distributed to Charlie but not Bob, and qudit 5 to Bob
but not Charlie, the information initially distributed in qudits 1,
2, and $a$ can also be concentrated to qudit 6 via the entangled
channel of Eq.~(\ref{channelg}). In this case, the procedure of RIC
is as follows. (S1) Alice, Charlie, and Bob perform GBMs on the
qudit-pairs ($3,a$), ($2,4$), and ($1,5$), respectively. (S2) Each
party tells Diana the measurement outcome by sending $2\log d$ bits
of classical information. (S3) Diana performs the conditional local
operation on qudit 6. This can be easily verified by
Eqs.~(\ref{clone2}), (\ref{channelg}), and (\ref{swapping}). A
schematic picture for this case is shown in Fig.~2. There are also
other cases of distribution of the entangled channel with which the
RIC can be achieved. However, if qudits 3 and 4 are simultaneously
distributed to Bob and Charlie (see, e.g., Fig.~3), RIC cannot be
achieved generally for $d>2$ by the same entangled channel of
Eq.~(\ref{channelg}) without special superposition coefficients as
shown later, which can be understood from Eq.~(\ref{clone3}). Note
that there is no such constraint for qubit-RIC, because the
inequality of Eq.~(\ref{clone3}) does not hold for $d=2$. Thus this
is a minor difference between qudit-RIC and qubit-RIC.

Equation (\ref{channelg}) contains a broad family of pure entangled
states. We now consider some special cases. Assuming $u=v=0$, $n'=c$
(an arbitrary nonnegative integer that is less than $d$), and
$C_{m'c}=1/\sqrt{d}$ for all $m'$, Eq.~(\ref{channelg}) reduces to
\begin{eqnarray}
\label{channels1}
|\Psi^{s_1}\rangle_{3456}&=&\frac{1}{\sqrt{d}}\sum\limits_{m'=0}^{d-1}|B^{m',c}\rangle_{34}|B^{\overline{-m'},\overline{-c}}\rangle_{56}\nonumber\\
 &=& \frac{1}{\sqrt{d}}\sum\limits_{j=0}^{d-1}|j\rangle_3|\overline{j+c}\rangle_4|j\rangle_5|\overline{j-c}\rangle_6,
\end{eqnarray}
i.e., a multilevel (or generalized) Greenberger-Horne-Zeilinger
(GHZ) state \cite{58AJP1131,63PRA022104}. In this case, there is no
constraint on the channel distribution, i.e., qudits 3, 4, and 5 can
be arbitrarily distributed to Alice, Bob, and Charlie, each party
one qudit. Assuming $u=v=0$, $C_{00}=\alpha$, $C_{m'0}=\beta$
($m'=1,2,\cdots,d-1$), and $C_{m'n'}=\gamma$ ($m'=0,1,\cdots,d-1$;
$n'=1,2,\cdots,d-1$), the entangled channel of Eq.~(\ref{channelg})
reduces to
\begin{eqnarray}
\label{channels2}
|\Psi^{s_2}\rangle_{3456}&=&\alpha|B^{0,0}\rangle_{34}|B^{0,0}\rangle_{56}
  +\beta\sum\limits_{m'=1}^{d-1}|B^{m',0}\rangle_{34}|B^{\overline{-m'},0}\rangle_{56}\nonumber\\
  &&+\gamma\sum\limits_{m'=0,n'=1}^{d-1}|B^{m',n'}\rangle_{34}|B^{\overline{-m'},\overline{-n'}}\rangle_{56}.
\end{eqnarray}
For the case $d=2$, it can be proved that the state of
Eq.~(\ref{channels2}) is the same as that of
Eq.~(\ref{telecloning}). This indicates that the four-qubit
entangled state of Eq.~(\ref{telecloning}) can be competent for
implementing both telecloning and RIC, two inverse processes. In
other words, the aforementioned telecloning and RIC for $d=2$
(qubit) can be achieved by using the same entangled channel.
However, such a result is not applicable to $d>2$ (qudit). This is
another difference between qudit-RIC and qubit-RIC. According to
Ref.~\cite{59PRA156}, the states of Eqs.~(\ref{channels1}) and
(\ref{channels2}) with $d=2$ are not equivalent to each other, i.e.,
cannot be transformed into each other by LOCC. It can be verified
that the states of Eqs.~(\ref{channels1}) and (\ref{channels2}) with
$d>2$ are also LOCC inequivalent. This implies that Eq.~(11)
contains at least two inequivalent types of genuine four-partite
pure entangled states. In other words, different types of genuine
four-partite pure entangled states can implement a same multiparty
QIP task, three-to-one RIC. Such a phenomenon is counterintuitive,
since a given QIP task can be achieved by only typical structure of
entangled states and different types of entangled states are usually
competent for implementing different QIP tasks. It has been shown
\cite{74PRA062320,8QIP431} that quantum teleportation can be
deterministically implemented by using both multiqubit W and GHZ
states, two inequivalent genuine multiqubit entangled states
\cite{62PRA062314}. However, teleportation is a two-party
communication, and the W and GHZ states in fact play the same role
as the bipartite entangled state, i.e., only the bipartite
entanglement of them is exploited. In contrast, RIC is a multiparty
communication (each party holds one particle of the entangled
channel), and the states of Eqs.~(\ref{channels1}) and
(\ref{channels2}) play a role of multipartite entanglement.

We now show that the quantum channel of our RIC can also be a broad
family of mixed entangled states. Let
$C_{m'n'}=\delta_{m',M}\delta_{n',N}$, where $M$ and $N$ are two
arbitrarily chosen nonnegative integers that are less than $d$. Then
the quantum channel of Eq.~(\ref{channelg}) reduces to a product
state of two generalized Bell states,
\begin{equation}
\label{channels3}
  |\Psi^{s_3}\rangle_{3456}=|B^{M,N}\rangle_{34}|B^{\overline{u-M},\overline{v-N}}\rangle_{56}.
\end{equation}
Because the two constants $M$ and $N$ are arbitrary, we can deduce
that the quantum channel of our RIC can also be the following form
of mixed entangled states:
\begin{equation}
\label{channelrho}
 \rho_{3456}=\sum\limits_{m',n'=0}^{d-1}|C_{m'n'}|^2|B^{m',n'}\rangle_{34}\langle B^{m',n'}|
  \otimes|B^{\overline{u-m'},\overline{v-n'}}\rangle_{56}\langle B^{\overline{u-m'},\overline{v-n'}}|.
\end{equation}
This can be easily proved by resorting to a purified state of
$\rho_{3456}$,
\begin{equation}
 |\Psi^{\rho}\rangle_{3456XY}=\sum\limits_{m',n'=0}^{d-1}C_{m'n'}|B^{m',n'}\rangle_{34}
 |B^{\overline{u-m'},\overline{v-n'}}\rangle_{56}|B^{m',n'}\rangle_{XY}.
\end{equation}
Particularly, by carrying out the same procedure as before [see
Eqs.~(\ref{whole})-(\ref{teleportation})], the information of
$|\psi\rangle_{12a}$ can also be concentrated in qudit 6 via the
entangled channel $|\Psi^{\rho}\rangle_{3456XY}$. In the whole
process, qudits $X$ and $Y$ are not touched, and thus can be traced
out at any time. This finishes the proof that the mixed state
$\rho_{3456}$ can be competent for our RIC. For the case $d>2$, and
using the entangled channel $\rho_{3456}$ with $|C_{m'n'}|\neq 1/d$,
qudits 3 and 4 can also not be simultaneously distributed to Bob and
Charlie, otherwise, the information of $|\psi\rangle_{12a}$ cannot
be successfully concentrated to qudit 6. This can be understood from
Eq.~(\ref{clone3}) and that $\rho_{3456}$ with $|C_{m'n'}|\neq 1/d$
cannot be expanded as the same form as Eq.~(\ref{channelrho}) with
respect to the $2:2$ partition $\{\{3,5\},\{4,6\}\}$ or
$\{\{3,6\},\{4,5\}\}$. However, there is no such a constraint for
qubit-RIC \cite{86PRL352,73PRA012318}.

If we set $u=v=0$ and $|C_{m'n'}|=1/d$, Eq.~(\ref{channelrho})
reduces to
\begin{equation}
\label{channelrho'}
 \rho'_{3456}=\frac{1}{d^2}\sum\limits_{m',n'=0}^{d-1}|B^{m',n'}\rangle_{34}\langle B^{m',n'}|
  \otimes|B^{\overline{-m'},\overline{-n'}}\rangle_{56}\langle B^{\overline{-m'},\overline{-n'}}|.
\end{equation}
By Eq.~(\ref{swapping}), we can rewrite $\rho'_{3456}$ as
\begin{equation}
\label{channelrho''}
  \rho'_{3456}=\frac{1}{d^2}\sum\limits_{m',n'=0}^{d-1}|B^{m',n'}\rangle_{36}\langle B^{m',n'}|
  \otimes|B^{\overline{-m'},\overline{-n'}}\rangle_{54}\langle B^{\overline{-m'},\overline{-n'}}|.
\end{equation}
For $d=2$, $\rho'_{3456}$ is exactly the Smolin state
\cite{63PRA032306}, a four-qubit unlockable bound entangled state.
The Smolin state is fully symmetric; i.e., it is unchanged under
permutation of any two qubits. This leads to the Smolin state being
separable with respect to any $2:2$ partition of $\{3,4,5,6\}$. For
$d>2$, $\rho'_{3456}$ also describes an unlockable bound entangled
state. It can be seen from Eqs.~(\ref{channelrho'}) and
(\ref{channelrho''}) that for any two qudits $x\neq y \in
\{3,4,5,6\}$, there exists at least one partition $\{G_1,G_2\}$
($G_1\cap G_2=\emptyset$ and $G_1\cup G_2=\{3,4,5,6\}$) with $x\in
G_1$ and $y\in G_2$ such that $\rho'_{3456}$ is separable with
respect to this partition, which implies that it is impossible to
distill out pure entanglement between $x$ and $y$, even between
$G_1$ and $G_2$, by LOCC, as long as $G_1$ and $G_2$ remain
spatially separated. Thus $\rho'_{3456}$ is undistillable when the
four particles are spatially separated. The unlockability or
activability of $\rho'_{3456}$ is obvious. Particularly, it can be
unlocked as follows. Let qudits 3 and 4 (3 and 6) join together and
perform a GBM on them. Then depending on the measurement outcome
qudits 5 and 6 (4 and 5) is projected in a generalized Bell state.
That is, pure entanglement is distilled out between qudits 5 and 6
(4 and 5). However, $\rho'_{3456}$ with $d>2$ is an asymmetric but
not symmetric unlockable bound entangled state, because it is not
separable with respect to the $2:2$ partition $\{\{3,5\},\{4,6\}\}$.
In addition, it can be verified that $\rho'_{3456}$ cannot be
superactivated for $d>2$, which presents a striking contrast to the
Smolin bound entangled state being superactivable
\cite{90PRL107901,72PRA060303}. These results indicate that there
exists an analog to the Smolin state in multilevel systems; however,
it has some different features. Note that the asymmetric four-qudit
unlockable bound entangled state $\rho'_{3456}$ is not contained in
Ref.~\cite{75PRA052332}. Therefore, it is a ''new'' asymmetric
unlockable bound entangled state.

As shown above, many different types of entangled states, including
mixed entangled states as well as pure ones, can be exploited as the
quantum channel of three-to-one RIC. The pure states can be
double-Bell states and LOCC inequivalent genuine four-partite
entangled states. The mixed states can even be bound entangled
states. However, it can be verified that all these states have a
common feature that they have $d^2$ common commuting stabilizers
$\{S^{jk}=U^{\overline{-j},k}_3\otimes U^{j,k}_4\otimes
U^{\overline{-j},k}_5\otimes U^{j,k}_6: j,k=0,1,\cdots, d-1\}$. That
is, for any $j$ and $k$,
$\mathrm{tr}(S^{jk}|\Psi^g\rangle_{3456}\langle\Psi^g|)=\mathrm{tr}(S^{jk}\rho_{3456})=1$.

\section{discussion and conclusion}

We now give a brief discussion on the physical or experimental
realization of the RIC presented in Sec.~II. Light quantum states
can be utilized for implementing qudits by exploiting various
degrees of freedom of photons, such as polarization
\cite{71PRA062337,73PRA063810,77PRA015802}, orbital angular momentum
(OAM) \cite{412N313,3NP305}, path mode
\cite{55PRA2564,102PRL153902,83PRA062333}, time bin
\cite{69PRA050304}, or a combination of different degrees of freedom
(see, e.g., \cite{80PRA022326,81PRA052317}), and so on. In deed,
many optical realizations, manipulations, and applications of qudits
and entangled qudits with the aforementioned degrees of freedom have
been experimentally demonstrated
\cite{412N313,102PRL153902,69PRA050304,81PRA052317,92PRL167903,91PRL227902,93PRL053601,100PRL060504,97PRL023602}.
As to the experimental implementation of RIC for qudits, one mainly
needs to consider three points as follows: (i) preparation of the
entangled channel, i.e., preparing $d$-level Bell states (bipartite
maximally entangled states) or GHZ states, or the unlockable bound
entangled states of Eq.~(\ref{channelrho'}); (ii) realization of
$1\rightarrow 2$ optimal telecloning (or cloning) of a $d$-level
arbitrary quantum state; (iii) implementation of GBM in $d$-level
systems. All these building blocks are achievable in quantum optics
as illustrated below. Many schemes for generating high-dimensional
entangled states of photonic qudits have been proposed and
demonstrated. Experimental realization of two-qutrit ($d=3$)
maximally entangled states (generalized Bell-basis states, or can be
transformed into any one of $d^2$ Bell-basis states by local
operations, \emph{to be uniformly referred to as generalized or
qudit Bell states}) with each qutrit encoded by three polarization
states of two frequency-degenerate photons in the same
spatiotemporal mode (biphoton) has already been reported
\cite{88PRL030401,75PRA022325}. A flexible scheme for generating
various entangled states (including generalized Bell states) of two
ququarts ($d=4$) using polarization degrees of freedom of the
frequency-nondegenerate biphoton was put forward \cite{75PRA034309},
which is scalable to generating various multiququart entangled
states. Simple schemes for creating $h$-color entangled states
(including generalized Bell states or GHZ states) of $N$ qudits
($1\leqslant h\leqslant N$) with multiphoton polarization were also
proposed \cite{77PRA015802}, in which $N$ and the dimension $d$ can
be arbitrarily large with sacrifice of success probability in
principle. By using OAM of photons, the Zeilinger research group and
co-workers realized qutrit Bell states of two photons with different
methods \cite{91PRL227902, 8NJP75}, and also showed that two-qudit
photonic entanglement up to $d=21$ are experimentally realizable via
a spatial light modulator \cite{90APL261114}; Torres \emph{et al.}
presented another method to generate two-photon high-dimensional
maximally entangled states and demonstrated the preparation of nine
Bell-basis states of two qutrits, which is based on the use of a
coherent and engineerable superposition of modes as a pump signal
\cite{67PRA052313}; these methods together with OAM beam splitter
\cite{71PRA042324} make it possible to create multi-qudit entangled
states, e.g., multilevel GHZ states. Four- and eight-level Bell
states of two photons with path-mode have recently been reported
\cite{102PRL153902,94PRL100501}; we conjecture that these techniques
together with $2d$-port beam splitter \cite{55PRA2564} could be used
to create $d$-level GHZ states, as a natural extension of $2\times
2$-port beam splitter synthesizing qubit GHZ states from qubit Bell
states. Energy-time or time-bin generalized Bell states of two
photonic qutrits have also been experimentally realized
\cite{93PRL010503}. The $d$-level unlockable bound entangled state
of Eq.~(\ref{channelrho'}) can be created from two identical
$d$-level Bell-basis states by randomly (with equal probability) and
simultaneously performing the pairwise operations
$\{U^{m,n},U^{\overline{-m},\overline{-n}}\}$ on two qudits
belonging to, respectively, different Bell pairs \cite{5NP748}.
Recently, a flexible scheme for $1\rightarrow 2$ optimal universal
cloning of a photonic ququart has been proposed and experimentally
demonstrated by Nagali \emph{et al.} \cite{105PRL073602}, which is
generally applicable to quantum states of arbitrarily high dimension
and is scalable to an arbitrary number of copies
\cite{105PRL073602,3NPh720}. As to the optical implementation of
GBM, two schemes have also been put forward. Halevy \emph{et al.}
proposed and experimentally demonstrated a realization of
three-level GBM, with each qutrit being represented by the
polarization of biphoton \cite{106PRL130502}. Du\v{s}ek presented a
method to implement GBM of path-mode-encoded qudits \cite{199OC161}.
The aforementioned schemes of cloning and GBM could also be
generalized or applied to other optical systems mentioned above
because of the permission of mapping or converting between different
degrees of freedom \cite{80PRA062312,103PRL013601}. The
illustrations and analysis given above appear possible for
experimental implementation of RIC in multilevel systems.

In conclusion, we have studied the RIC in multilevel systems, and
shown that the information of the three-qudit universal cloning
state can be remotely and deterministically concentrated to a single
qudit via an entangled channel with LOCC. Minor differences of
qudit-RIC with qubit-RIC have also been analyzed. It has been shown
that there are minor constraints on the distribution of the general
entangled channel for qudit-RIC, but none for qubit-RIC. Moreover,
telecloning and RIC for qubits can be achieved by using the same
entangled channel, but there is no such a feature for qudits.

We investigated many types of entangled states as the quantum
channel, including mixed entangled states as well as pure ones, and
found some interesting phenomena. Similar to qubit-RIC, qudit-RIC
can also be implemented by an unlockable bound entangled state.
Though such a multilevel bound entangled state has a similar form to
the Smolin bound entangled state, it has some different features. As
a matter of fact, they belong to different types of unlockable bound
entangled states: the former one is asymmetric and the latter one is
symmetric. It has been shown that the quantum channel of RIC can be
different types of genuine four-partite pure entangled states which
are LOCC inequivalent. Moreover, we found that all these states,
which can act as the quantum channel of RIC, have $d^2$ common
commuting stabilizers. This implies that the states which have
common stabilizers could be competent for implementing
(deterministically) some same QIP tasks. We hope these phenomena
will stimulate more research into the topic of dividing or
classifying entangled states by the usefulness for typical QIP
tasks. Maybe this needs resorting to the stabilizers. Then the
\emph{genuine} multipartite pure entangled states which can be
competent for implementing (deterministically) one or more same
\emph{multiparty} tasks may be LOCC inequivalent. In view of the
fact that entanglement is a very important physical \emph{resource}
for QIP, this topic should be meaningful and important.

\section*{Acknowledgements}
This work was supported by the National Natural Science Foundation
of China (Grant Nos. 11004050 and 11174040), the Key Project of the
Chinese Ministry of Education (Grant No. 211119), the Scientific
Research Fund of the Hunan Provincial Education Department of China
(Grant Nos. 09A013 and 10B013), the Science and Technology Research
Foundation of Hunan Province of China (Grant No. 2010FJ4120), and
the Excellent Talents Program of Hengyang Normal University of China
(Grant No. 2010YCJH01).

\newpage

\begin{figure}
  \center
  \includegraphics[width=7cm,height=5cm]{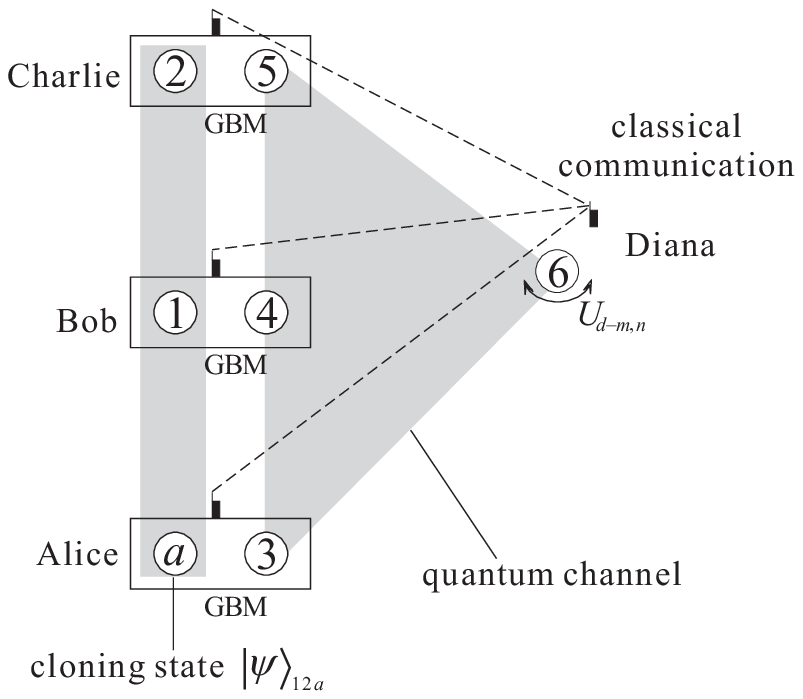}
  \caption{Schematic picture showing the successful concentration of
information from Alice, Bob, and Charlie at the remote receiver,
Diana, in the case in which qudits 3, 4, and 5 of the four-qudit
entangled sate acting as the quantum channel are distributed to
Alice, Bob, and Charlie, respectively.}
\end{figure}

\vspace{5pt}

\begin{figure}
   \center
   \includegraphics[width=7cm,height=5cm]{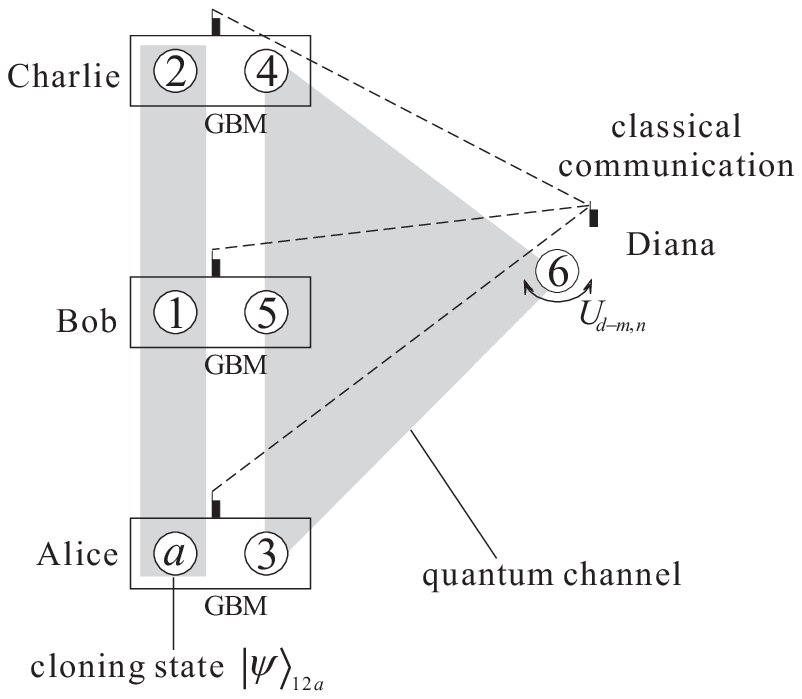}
   \caption{Schematic picture showing the successful concentration of
information from Alice, Bob, and Charlie at the remote receiver,
Diana, in the case in which qudits 3, 4, and 5 of the four-qudit
entangled state acting as the quantum channel are distributed to
Alice, Charlie, and Bob, respectively.}
\end{figure}

\vspace{5pt}

\begin{figure}
  \center
  \includegraphics[width=7cm,height=5cm]{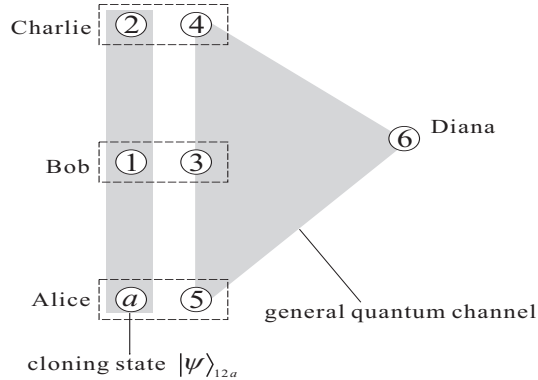}
  \caption{Schematic picture showing the failure of concentrating
information from Alice, Bob, and Charlie to the remote receiver,
Diana, using the general entangled channel $|\Psi^g\rangle_{3456}$
[see Eq.~(\ref{channelg})] or $\rho_{3456}$ [see
Eq.~(\ref{channelrho})], in the case in which qudits 3, 4, and 5 are
distributed to Bob, Charlie, and Alice, respectively.}
\end{figure}

\end{document}